\documentclass[preprint,preprintnumbers,amsmath,amssymb,elsart]{revtex4}
\usepackage{graphicx,psfrag,bbm,latexsym,color,dcolumn,bm,dsfont,bbm,color,mathrsfs,latexsym,amsmath,amsfonts,
amssymb,epsfig,natbib}

\newcommand{\beq}
{\begin{equation}}

\newcommand{\eeq}
{\end{equation}}

\newcommand{\beqa}
{\begin{eqnarray}}

\newcommand{\eeqa}
{\end{eqnarray}}

\newcommand{\bw}
{\begin{widetext}}

\newcommand{\ew}
{\end{widetext}}

\begin{document}

\title{The Decoherence of the Electron Spin Polarization and Meta-stability of ${}^{13}C$ Nuclear Spins in Diamond}
\author{P.R. Crompton}
\affiliation{Department of Applied Maths, School of Mathematics, University of Leeds, Leeds, 
LS2 9JT, UK.}
\vspace{0.2in}
\date{\today}

\begin{abstract}
\vspace{0.2in}
{Following the recent successful manipulation of entangled ${}^{13}C$ atoms on the surface of Diamond we 
calculate the decoherence of the electron spin polarization in Diamond via a nonperturbative treatment of the time-dependent Greens function of the Central-Spin model, describing the Hyperfine couplings of the electron to a bath of interacting ${}^{13}C$ atoms, for arbitrary initial polarizations, applied field strengths, and for up to eight entangled ${}^{13}C$ atoms. We compare our numerical results, which are regularized by $\zeta$-function prescription, with the exact treatment available that is available in the fully initially polarized limit of the non-Markovian dynamics regime, and comment on the relevance of dangerously irrelevant ${}^{13}C$ dipole-dipole interactions to the meta-stability of ${}^{13}C$ nuclei flip-flop processes.}

\vspace{0.1in}
\end{abstract}

\maketitle


Recently, several experimental groups \cite{9}-\cite{34} have investigated the feasibility of processing Quantum Information via the manipulation of optically excited electron spins \cite{58} in Diamond. These studies focus on the Diamonds' optically active Nitrogen Vacancy centers and the triplet that is formed there through the Zeeman splitting of a Nitrogen impurity coupled to a lattice vacancy \cite{61}. There are two known charged states of this point defect, which are neutral and negatively charged, respectively, with spin states of $m_{s}=0$ and $m_{s}=1$. Hence, stable qubits have beed formed in Diamond consisting of a single electron spin in the $m_{s}=1$ state coupled to a bath of the neighbouring nuclear spins which slowly decoheres via the spectral diffusion of the nuclear spin polarizations.  Diamond offers several advantages over conventional Quantum Dot materials, such as GaAs for this purpose \cite{12}\cite{13}. This is  because the density of states of this system is relatively low (with a low electron-phonon coupling), whilst the oscillator strength for the electron dipole transition is relatively large. This is true in particular for the electron's coupling to the ${}^{13}C$ atoms which have a natural abundance of 1.1\% \cite{32}. Consequently, it has been reported that electron spin-dephasing times of $0.5 - 6 \mu s$ can be achieved in Diamond, at 300K - without the need for cryogenic cooling \cite{9}-\cite{34}, which makes Diamond an attractive candidate for Quantum Information Processing.

This spin decoherence of single electrons in the $m_{s}=1$ state of the Nitrogen Vacancy centers of Diamond can thus be modelled via the competition between; the spin-orbit coupling of the electron and ${}^{13}C$ nuclei, the Hyperfine coupling between the electron and a spin-bath of ${}^{13}C$ nuclear spins, and the dipole-dipole interaction between the ${}^{13}C$ bath nuclei \cite{1}, which together is given via the following Central-Spin model Hamiltonian \cite{61}\cite{2}-\cite{8}, 

\begin{equation}
\label{central}
H = \epsilon_{S} {\bm{S}}^z  + 
    \sum_{i=1}^{N} \epsilon_{I} {\bm{I}}_{i}^{z} +
    \sum_{i=1}^{N} A_{i} {\bm{S}}.{\bm{I}}_{i}^{z} + 
    \sum_{i=1}^{N} \sum_{i\neq j,\, j=1}^{N} \frac{\epsilon_{dd}}{r^{3}_{ij}} 
    \left[ \left( {\bm{I}}_{i}.{\bm{n}}_{i j} \right ) 
           \left( {\bm{I}}_{j}.{\bm{n}}_{i j} \right )  
          -{\bm{I}}_{i}.{\bm{I}}_{j} \right]   
\end{equation} 

where ${\bm{S}}^z ({\bm{I}}_{i}^{z})$ is an operator for the longitudinal spin of the electron in the adjoint representation ($i$th nuclear spin), ${\bm{I}}_{i}$ is an operator for the $i$th ${}^{13}C$ nucleus in the spin-bath, $N$ is the number of bath nuclei and ${\bm{n}}_{i j}$ is a unit vector linking the centers of the $i$th and $j$th ${}^{13}C$ nuclear dipole moments with separation $r_{ij}$. Since the relative strength of the couplings of the Zeeman, Hyperfine and dipole-dipole interactions in Diamond are $\{ \epsilon_{S} = 115.6 \,\mu eV T^{-1}, \epsilon_{I} = 0.1196 \,\mu eV T^{-1}, A_{i} = 0.09873 \,\mu eV, \epsilon_{dd} = 1.2566\times 10^{-4} \,\mu eV \,\}$\cite{2}, naively, the dipole-dipole ${}^{13}C$ interactions can be neglected. However, what makes this model very interesting is the difference between the $m_{s}=0$ and $m_{s}=1$ states in the applied field, since for $m_{s}=0$ the nuclear spins all precess with the same frequency, whilst for $m_{s}=1$ the bath nuclei spins diffuse via the Hyperfine interaction \cite{9}. Hence, although we choose ${\bm{S}}^z$ to represent the triplet state of the electron via the adjoint representation, if the field strength is tuned to the level at which the energies of the $m_{s}=0$ and $m_{s}=1$ states become equal ($\sim 500G$), the otherwise irrelevant ${}^{13}C$ dipole-dipole can become important to the decoherence of the electron spin through quantum tunnelling between the triplet and singlet state. In such a case the Hyperfine intercation of the electron-dipole interaction should then lead to a metastability in the ${}^{13}C$ nuclei spin polarizations.

Via Sturm-Liouville theory (in the absence of the dipole-dipole interaction), the longitudinal and transverse spin components of a single electron in the Central-Spin model are exactly separable, which yields the following Nakajima-Zwanzig generalized master equations (GMEs) \cite{7}, 

\begin{equation}
\langle {\dot{{\bm{S}}^{z}}} \rangle_{t} 
= N_{z}(t) -i\int_{0}^{t} dt' 
\, \Sigma_{zz} (t-t') \langle {\bm{S}}^{z} \rangle_{t'},
\end{equation}
\begin{equation}
\langle \dot{{\bm{S}}^{+}} \rangle_{t} = i\omega_{n} \langle {\bm{S}}^{+} \rangle_{t} 
-i\int_{0}^{t} dt' \, \Sigma_{++} (t-t')\langle {\bm{S}}^{+}\rangle_{t'}
\end{equation}

where $N_{z}, \Sigma_{zz}$ and $\Sigma_{++}$ are given by matrix elements of the reduced self energy $\Sigma(t-t')$, $\omega_{n} \equiv B(\epsilon_{S} - \epsilon_{I}) + h_{I}$, with $h_{I}$ an eigenvalue of the initial angular momentum eigenstates of the bath, and $B$ is the applied field strength. From which, the time dependence of the electron spin polarization can be evaluated via Laplace transform. Importantly, even for this simple case, the above GMEs are nonperturbative for  $\omega_{n}/N > 1$ and so the self-energy contributions do not resum and grow as a function of time. To make this GME analytically tractable in the pertubative regime (for arbitrary initial polarizations in the high field limit), the above self energies can be reparameterised as continuous functions of the Hyperfine couplings (via $A_{k} \rightarrow x$). In the Born approximation (with $\Sigma_{zz} = \Sigma_{zz}^{(2)}(I_{\pm}(s))$) this yields the functions,  

\begin{equation}
I_{\pm}(s) = \frac{1}{4N} \sum_{k} \frac{A^{2}_{k}}{s\mp \frac{iA_{k}}{2}} \quad \rightarrow \quad 
I_{\pm}(s) = \frac{d}{m} \int_{0}^{1} dx \frac{x |\ln x|^{\nu}}{s \mp ix}, \quad \nu = \frac{d}{m} -1, 
\end{equation}

where $P_{\bm{I}}(m)$ is the probability of finding a nuclear spin $\bm{I}$ with z-projection $m$, and $d$ the dimension of the system \cite{7}. Hence, for $\nu < 1$ the electron spin decoherence envelope is oscillatory and is governed by nuclei close to the central electron (Markovian), whereas for $\nu > 1$, the electron spin decoherence envelope is slowly decaying and governed by the nuclei farthest away (non-Markovian). However, in the field splitting problem we consider in (1) we need to take a more general approach to evaluating these integrals, since delineating the regimes via $\nu$ is adversely affected by the quantum tunnelling, as the Markovian and non-Markovian regimes can become interchanged \cite{70}. In a renormalization group sense, $\nu$ is a dangerously irrelevant scaling parameter through the flip-flop dynamics of the ${}^{13}C$ nuclei \cite{27}\cite{71}\cite{50}.

To tackle the field splitting case in (1) we follow the recent approach of using fractional derivatives \cite{53}\cite{54}\cite{52}, to generalise the Sturm-Liouville theory derivation of the GME in (2) and (3). This approach is based on the generalisation of the Brownian motion of the paths that are required for Euclidean Feymann path integrals in quantum mechanics to L\'{e}vy random processes \cite{52}. Rather than the GME being meromorphically continued into poles, in this case, the GME is meromorphically continued onto a annulus, corresponding to the entangled nuclear spin states \cite{50}. Formally, the L\'{e}vy path integral generates a functional measure in the space of either left or right continued functions, and has only discontinuities of the first kind (first order derivatives) corresponding to whichever spin-flip state of the ${}^{13}C$ nuclei is the most energetically favourable. We therefore replace the time derivatives in (2) and (3) by the following generalised (Caputo) fractional time derivative \cite{51}\cite{21},

\begin{equation}
D_{0\,\,*t}^{\nu} \left( {\bm{S}}^{- }(t) {\bm{S}}^{+}(0) \right) = \frac{1}{\Gamma(1-\nu)} \int_{0}^{t} dt'
\frac{1}{(t-t')^\nu} \frac{\partial}{\partial t'} \left( {\bm{S}}^{- }(t') {\bm{S}}^{+}(0^{+}) \right) 
\end{equation}

where ${\bm{S}}^{- }(t)$ and ${\bm{S}}^{+}(0)$ are time evolution operators for the electron spin (in the adjoint representation) at time $t=0$ and $t=t$ \cite{75}. The relevance of this particular product of evolution operators is that, following \cite{61}\cite{1}\cite{2}, the density matrix for the transverse electron spin in (1) is defined via, ${\rm{Tr}} \left[\, {\bm{S}}^{-}(t) {\bm{S}}^{+}(0^{+}){\bm{S}}^{+\dagger}(t) {\bm{S}}^{-\dagger}(0^{+}) \right ]$. Practically evaluating the Greens function propagators of this fractional derivative formalism then leads to overall constant in front the propagtors, compared to the conventional GME, since the Fourier transform of a fractional derivative is given by \cite{76},  

\begin{equation}
{\mathcal{F}} \left[ D^{\nu}_{\theta} \left( {\bm{S}}^{- }(t) {\bm{S}}^{+}(0) \right) \right] = -\psi^{\theta}_{\nu}(t) {\mathcal{F}} \left[ {\bm{S}}^{- }(t) {\bm{S}}^{+}(0) \right], \quad \psi^{\theta}_{\nu}(t) = |t|^{\nu} e^{i \,{\rm{sgn}} (t) \theta\pi/2} 
\end{equation}

The fractional derivative and adjoint prescription we make formally defines a regularization of electron spin decoherence via a $\zeta$-function prescription \cite{50}\cite{54}\cite{18}.
Our analysis of the decoherence of the electron spin polarization of Nitrogen Vacancy centers of Diamond extends previous analyses in two ways; firstly, we do not drop any off-diagonal (dipole-dipole) interaction terms in order to make the analysis more suitable for the GME approach in \cite{7}, and secondly, we use the adjoint representation to explicitly calculate the nucleation of the of ${}^{13}C$ flip-flop dynamics via the factor $|t|^{\nu}$.

\section{Discrete Fourier Transform}

In the nonperturbative regime, the electron spin decoherence can be calculated for the GME in (1) using a range of numerical schemes \cite{6}; exact diagonalization \cite{8}\cite{56}\cite{57}, loop-cluster expansion \cite{1}\cite{2} and via a discrete Fourier transform. In the following analysis we use latter method (although the $\zeta$-function prescription can also be applied numerically via \cite{50}) since we are able to keep the long-time cutoff fixed in the discrete Fourier Transform via the lattice spacing, which allows us to make an accurate comparison of $|t|^{\nu}$ for different field strengths and polarizations. The advantage of this approach is in having in a proper regularization of the GME in a treatment which also includes a full dipole-dipole interaction and the field splitting of the Nitrogen Vacancy center.

We calculate the decoherence of the transverse spin electron spin by using a discrete Fourier transform to calculate time-dependent Greens function of the Hamiltonian in (1), for up to a maximum of $N=8$ nuclei. Although $N=8$ is an experimentally relevant number for current entanglement studies \cite{32}\cite{33}\cite{12}, this small bath size can further justified for larger quantum dot systems via the convergence of Chebyshev polynomials \cite{6}. Explicitly: if the state vector of the spin-bath is represented via, 

\begin{equation}
| \chi \rangle = \sum_{k=1}^{N_{B}} \alpha_{i} | \, \beta_{1} \, \beta_{2} \, ... \, \beta_{\, N_{B}-1} \, \rangle, \quad 
\sum_{k=1}^{N_{B}} |\alpha_{k}|^2 = 1  
\end{equation}

where $\beta$ is either 0 or 1 (depending on whether the $i$th bath nuclei is spin up or spin down), $\alpha_{k}$ is 
a random variable, and there are $N_{B} = 2^N$ bath states, then, since the off-diagonal elements are of order 
$2^{-N_{B}}/\sqrt{N_{B}}$, agreement with the electron spin polarization to 6\% of the value for the full system (of $N \sim 10^6$) can be obtained from sampling clusters of just $N=8$ nuclei. Furthermore, this limit is an exact bound for the random L\'{e}vy random processes that form the (canonical) basis of our approach \cite{55}.

We obtain the energy-dependent Greens function of the Central spin model for Diamond via an exact inversion of the 
Hamiltonian in (1) using Gauss-Jordan elimination,

\begin{equation}
\label{G-J}
[G(\varepsilon )]_{i,j} = [\, (\varepsilon - H)^{-1} \,]_{i,j}
\end{equation}

where $G(\varepsilon)$ is then a $2^{2N+2}$ matrix, defined in the product basis $|\psi\rangle = |
S\otimes I_{1} \otimes I_{2} ... \otimes I_{N} \rangle$. We repeat this calculation for $M$ discrete values of energy, 
$\varepsilon$, where $G_{n}(\varepsilon) = G(\varepsilon)|_{\varepsilon = n \Delta\epsilon}$ with 
$\Delta\varepsilon = E/M$ and $E$ the finite energy domain of the system, which we choose to be of $\mathcal{O}(\epsilon_{S})$. The time-dependent Greens function of the Hamiltonian in (1) is then given by the following discrete Fourier transform, $G_{k}(t) = G(t)|_{t = k \Delta\tau}$, which we evaluated using the numerically efficient Four-Step Fast Fourier Transform \cite{15}, 

\begin{eqnarray}
\label{FT}
G_{k\,=\, qP + Q}(t) & = & \sum_{r=1}^{p} \left( e^{2\pi i \, rq/p} \left( e^{2\pi i \, rQ/pP} 
\sum_{R=1}^{P} G_{RP+r}(\varepsilon) e^{2\pi i \, RQ/P} \right)\right), \\ && \quad 0<r,q<p, \quad 0<R,Q<P, \quad pP=M.
\end{eqnarray}

where $\Delta\tau = T/M$ with $T$ the finite time domain of the system, and where the short range cutoff set via $E T = \hbar$ and $M=2^{14}$, \cite{4}. The time-dependence of the transverse spin of the electon at the Nitrogen Vacancy centers in Diamond is then given by the expectation,

\begin{equation}
\langle {\bm{S}}^{+} \rangle_{t} = \langle \, \psi^{\dagger}_{0} \, G^{\dagger}(t) \, {\bm{S}}^{z} \, 
G(t) \, \psi_{0} \rangle = \Theta(t) \langle \psi_{0} | \, {\bm{S}}^{-}(t) {\bm{S}}^{+}(0^{+}) \, | \psi_{0} \rangle 
\end{equation} 

where the initial electron state is defined to be $\psi_{0} = \left( \, |0\rangle + |1\rangle \, \right)/\sqrt{2}$ perpendicular to the initial direction of the nuclear spins,  \cite{61}\cite{1}\cite{2}. Note that our choice of adjoint representation and fractional derivative approach means that the above expectation for the transverse electron spin is real (and negative) and is also not normalised to unity. This allows us to quantify the nucleation of the of ${}^{13}C$ flip-flop dynamics via the factor $|t|^{\nu}$ in (6).

\begin{figure}
\epsfig{file=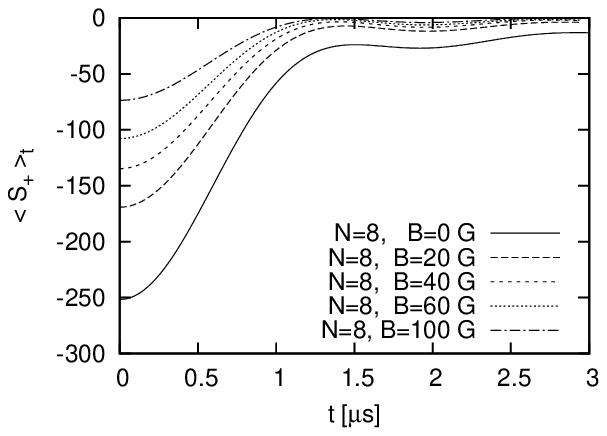, width=8cm}\epsfig{file=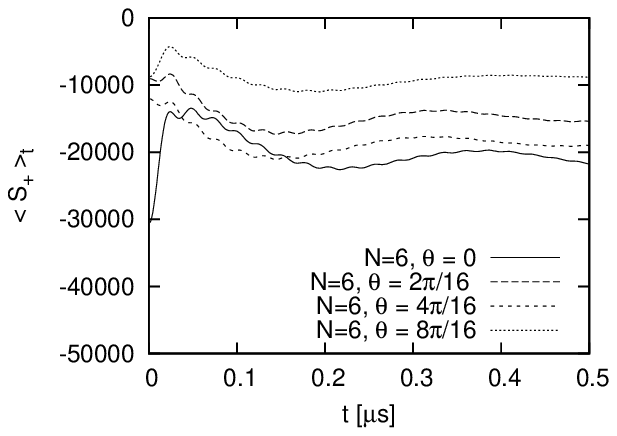, width=8cm}
\caption{Time dependence of the transverse component of electron spin, $\langle S^{+} \rangle_{t}$, for spin baths 
of $N=8$ and $N=6$ ${}^{13}C$ atoms in Diamond, respectively. For $N=8$ the applied field is varied between $0-100 G$, 
whereas for $N=6$ the applied field direction is varied relative to the $z$-axis at $100 G$.}
\end{figure}

\begin{figure}
\epsfig{file=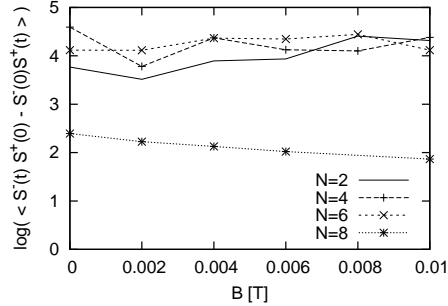, width=6cm}
\caption{Applied field strength dependence of the logarithm of the argument of the Caputo integral in (11), for spin baths of $N=2 - 8\,\,$ ${}^{13}C$ atoms in Diamond.}
\end{figure}

We compare our numerical results in the first instance with the exact results for the Central Spin model obtained in \cite{7} for the limits of large applied external magnetic field, and a fully polarized initial system. The zeroth order dynamics result for the transverse electron spin in (3) is given by, 

\begin{equation}
\label{zero}
\langle S_{+} \rangle_{t} = \langle S_{+} \rangle_{0} \exp[-t^{2}/2\tau_{c}^{2} + i(B(\epsilon_{S} - \epsilon_{I})+pN)t],  
\quad \tau_{c} = \frac{1}{\sqrt{N(1-p^{2})}}
\end{equation}

where $p$ is the fraction of initially polarized states, and $\tau_{c}$ the transverse electron decoherence time, which, from the perturbative expansion of (3), we can expect to be relevant to order $2(k+1)$ when the self-energy is supressed by a factor of $(\omega_{n}/N)^{k}$. From the second term in the exponent in (\ref{zero}), it follows that the decoherence time, $\tau_{c}$, becomes independent of the applied field strength for large values of field, and that the nonperturbative regime is therefore delineated by, $N |B_{z}| \gg |\frac{A}{g\mu_{B}}| \sim \mathcal{O}(10 \, \rm{Gauss})$. 

In Figure 1 we plot our evaluation of the time dependence of the transverse spin component of the electron, $\langle S_{+} \rangle_{t}$, for spin baths of $N=8$ and $N=6\,\,$ ${}^{13}C$ nuclei in Diamond. In the left plot (for $N=8$ nuclei) the applied field is varied between $0-100 \, \rm{Gauss}$, and in the right plot (for $N=6$ nuclei at $100 \, {\rm Gauss}$) the angle between the applied field and $z$-component of electron spin is varied between $\theta=0 - \pi/2$. Following our broad expectations from the zeroth order result in (12), the decoherence time increases in the left plot from $1 \mu s$ to $7 \mu s$ as a function of decreasing applied field strength, and increases in the right plot as a function of increasing (polarization), $\theta$, from $2 \mu s$ to $6 \mu s$. Note, that whilst for $N=8$ the expectation value of the transverse spin component of the electron tends to zero, $\langle S_{+} \rangle_{t\rightarrow \infty} = 0$, where we are closer to the above nonperturbative bound for $N=6$ the expectation tends to a nonzero value, $\langle S_{+} \rangle_{t\rightarrow \infty} \neq 0$. 

In Figure 2 we plot the applied field dependence of the logarithm of expectation value of the transverse spin component of the electron $\log \left( \langle S_{+} \rangle_{T} \right)$ which from (6) gives a measure the relative differences in $|t|^{\nu}$ as a function of applied field strength. From (4), near the upper bound of the memory kernel ($\nu > 1$) we expect a power law decay, where the electron spin decoherence envelope is slowly decaying and governed by the nuclei furthest from the central electron. As in Figure 1, evidently the systems for $N=2-6$ which are closer to the nonperturbative bound have larger associated values of $\nu$, which is also relatively insensitive to applied field strength. This is in contrast with the $N=8$ system that is furthest from the nonperturbative bound. This implies that the meta-stability of the ${}^{13}C$ nuclei (which is responsible for the value of $\nu$) decreases as a function of increasing applied field for larger bath sizes, as would be expected from (12).

\section{summary}

We have calculated the transverse electron spin decoherence for Nitrogen Vacancy center in Diamond, modelled as a Central-Spin electron spin in the adjoint representation coupled to a bath of up to eight entangled ${}^{13}C$ nuclei, with an additional dipole-dipole interaction also included between these bath nuclei to enable us to quantify the flip-flop dynamics of the bath. Our approach is based on a $\zeta$-function regularization of the quantum tunnelling bound between the triplet and singlet states of the Nitrogen Vacancy center, which we have introduced via fractional 
time derivatives \cite{51}\cite{21}. Our approach allows us to extend previous analyses in the nonperturbative regime, which we have probed via the numerical Four-Step Fast Fourier Transform by calculating the decoherence of the transverse electron spin in (1) via a discrete evaluation of the time-dependent Greens function of the Hamiltonian. Our results for the decoherence time of the transverse electron spin are consistent with recent experimental measurements for (proximal) spin baths of this size \cite{9}-\cite{34}\cite{12}, and furthermore, our investigation of the infrared cutoff dependence and regularization of our approach indicates that the stability of our analysis improves as a function of increasing bath size, as more channels become available for the quantum tunnelling.\\

Our thanks to V.I. Fal'ko for useful discussions.

\end{document}